# An algorithm based on a Cable-Nernst Planck model predicting synaptic activity throughout the dendritic arbor with micron specificity

Claire Guerrier[1,2][†][*], Tristan Dellazizzo Toth[3][†], Nicolas Galtier[4] and Kurt Haas[3].

[1]Université Côte d'azur, LJAD, CNRS UMR7351 - Nice, France.
[2]CNRS - IRL3457, CRM, Université de Montréal - Montréal, Canada.
[3]Center for Brain Health, UBC - Vancouver, Canada
[4]Independent researcher
[†]Contributed equally to this work: C. Guerrier and T. Dellazizzo Toth

[*]Corresponding author name and email address: Claire Guerrier claire.guerrier@univ-cotedazur.fr

## Abstract

Recent technological advances have enabled the recording of neurons in intact circuits with a high spatial and temporal resolution, creating the need for modeling with the same precision. In particular, the development of ultra-fast two-photon microscopy combined with fluorescence-based genetically-encoded $Ca^{2+}$-indicators allows capture of full-dendritic arbor and somatic responses associated with synaptic input and action potential output. The complexity of dendritic arbor structures and distributed patterns of activity over time results in the generation of incredibly rich 4D datasets that are challenging to analyze (Sakaki, 2020). Interpreting neural activity from fluorescence-based $Ca^{2+}$ biosensors is challenging due to non-linear interactions between several factors influencing intracellular calcium ion concentration and its binding to sensors, including the ionic dynamics driven by diffusion, electrical gradients and voltage-gated conductance.

To investigate those dynamics, we designed a model based on a Cable-like equation coupled to the Nernst-Planck equations for ionic fluxes in electrolytes. We employ this model to simulate signal propagation and ionic electrodiffusion across a dendritic arbor. Using these simulation results, we then designed an algorithm to detect synapses from $Ca^{2+}$ imaging datasets. We finally apply this algorithm to experimental $Ca^{2+}$-indicator datasets from neurons expressing jGCaMP7s (Dana *et al.*, 2019), using full-dendritic arbor sampling *in vivo* in the *Xenopus laevis* optic tectum using fast random-access two-photon microscopy.

Our model reproduces the dynamics of visual stimulus-evoked jGCaMP7s-mediated calcium signals observed experimentally, and the resulting algorithm allows prediction of the location of synapses across the dendritic arbor.

Our study provides a way to predict synaptic activity and location on dendritic arbors, from fluorescence data in the full dendritic arbor of a neuron recorded in the intact and awake developing vertebrate brain.

**keywords:** Calcium dynamics, synaptic organization, electrodiffusion, numerical simulations, mathematical modeling, python, PDE

# Background

A leading question in neuroscience is how neurons process synaptic inputs and generate an output encoded in action potential activity. In particular, it remains unanswered how the geometry of the dendritic arbor and distribution of the synapses contribute to the integration of synaptic information and the computation of the spike output. The historical model that dendrites serve to merely connect neurons and to passively convey information is currently challenged by recent findings demonstrating that the shape of the dendritic arbor, and by extension, the location of the synaptic inputs is key for non-linear signal integration (Redmond and Ghosh, 2005; Lavzin *et al.*, 2012; Gonzalez *et al.*, 2022). Indeed, dendritic patch-clamp electrophysiology and imaging of genetically encoded fluorescent $Ca^{2+}$-indicators have made recording activity within the dendritic arbor increasingly feasible, highlighting its involvement in information processing at a cellular level (London and Häusser, 2005; Basak and Narayanan, 2018). As a result, defining the role of the dendritic arbor and synaptic topography in information processing and encoding has become a leading area of research in developmental neuroscience. This field has taken a sizable leap forward by recent advances in ultra-fast multi-photon calcium imaging that provide unprecedented spatial and temporal resolutions of comprehensive dendritic calcium dynamics in intact and awake animals, but which also present novel challenges for their analysis (Sakaki *et al.*, 2020).

Relating the observed calcium dynamics, generated from full-dendritic arbor imaging of neurons expressing fluorescence-based calcium sensors, to underlying neural activity remains challenging. The regulation of intracellular calcium concentration by activity is complex, due to multiple sources of calcium and their nonlinear interactions. Calcium ions can enter neurons through synaptic and extrasynaptic glutamatergic receptors, and voltage-gated calcium channels, and can be released from intracellular endoplasmic reticulum stores. For studies focusing on tracking synaptic activity, there is a need for modeling to differentiate multiple calcium sources in order to determine the location and amplitude of synaptic responses. Such modeling requires understanding how voltage propagates throughout the dendritic arbor and its influence on ionic concentrations, especially in small neuronal compartments such as filopodia and dendritic spines (Holcman and Yuste, 2015; Savtchenko *et al.*, 2017).

Numerous models have been generated to characterize voltage dynamics in neurons. The standard approach consists of using Cable theory with Hodgkin-Huxley formalism (Qian and Sejnowski, 1989; Bower and Beeman, 1998; Carnevale and Hines, 2006), or integrate-and-fire type modeling (Keener *et al.,* 1981; Brette & Gerstner, 2005, Harkin *et al.*, 2022). Most of these models are geared toward neuronal network simulation, and do not include ionic dynamics. Other models are focused on ionic dynamics, and are based on the Nernst-Planck equation describing ionic fluxes, coupled to the Poisson equation, or to an electroneutral model (Lopreore *et al.*, 2008; Mory *et al.*, 2008; Xylouris *et al.*, 2010; Lu *et al.*, 2010; Pods *et al.*, 2013; Solbra *et al.*, 2018; Sætra *et al.*, 2020). The high non-linearity of the Poisson-Nernst Planck system of equations, and the presence of a thin boundary layer at the membrane renders such simulation and analysis a daunting task, especially when taking the complex geometries of dendritic arbors into consideration (Cartailler *et al.*, 2017; Savtchenko *et al.*, 2017).

Here, our goal is to realize fast simulations of voltage and ionic dynamics in detailed dendritic arbor geometries. To achieve this, we developed a model and simulations for voltage propagation and ionic electrodiffusion in the dendritic arbor. The model is based on a coupling between the Nernst-Planck equations to represent ionic fluxes due to the electrodiffusion of ions (Kirby, 2010), and a Cable-like equation representing voltage dynamics. We demonstrate that under specific assumptions, we can reduce the total ionic flux to a simple resistive flux, and hence decouple the equations, while accurately keeping track of calcium dynamics. This decoupling simplifies the program and enables faster simulations. Simulations of the simplified model are performed using the *Sinaps* Python library the authors developed previously (Galtier and Guerrier, 2022).

To validate the model, we compared simulation results with data from *in vivo* two-photon calcium imaging experiments, using neurons expressing the genetically encoded fluorescent protein jGCaMP7s. We observed a substantial discrepancy between the temporal scales of the action potential and jGCaMP7s fluorescence dynamics: from the duration of the action potential, ranging from sub-millisecond to several milliseconds, to the entry of calcium through voltage-gated channels that takes several milliseconds and persists for tens of milliseconds, and finally to the fluorescence response of calcium sensors as observed in the experimental data, that last for multiple seconds. Importantly, we found that the locations of fluorescence dynamics across the arbor proximate to synapses are different from the dynamics distant from a synapse. This allows for the discrimination and localization of potential synaptic activity from fluorescence-based calcium data.

In this paper, we first describe our model coupling the Nernst-Planck equations and a Cable-like equation in the full dendritic tree, and show that this system can be decoupled, to speed-up simulations. Next, using the decoupled system, we simulated calcium and jGCaMP7s dynamics in a full neuronal geometry. From these simulated results, we infer that the calcium dynamics as reported by the fluorescence changes in genetically encoded calcium sensors differ depending on the distance of the sampled point from an active synapse. We then propose an algorithm to detect possible synaptic activity in fluorescent datasets and test it on experimental measurements.

## Methods

**Experimental protocol**
**Animal rearing conditions:** Albino *Xenopus laevis* tadpoles were reared in a room temperature container of 0.1x Steinberg's solution (1x Steinberg's in mM: 10 HEPES, 58NaCl, 0.67KCl, 0.34Ca(NO3)2, 0.83 MgSO4, pH 7.4). They were reared in a 12 hour day/night cycle. All experimental procedures and housing conditions were approved by the University of British Columbia
Animal Care Committee and were in accordance with the Canadian Council on Animal Care (CCAC) guidelines.

**Expression of genetically-encoded fluorophores:** Single-cell electroporation (Haas *et al.*, 2001) was employed to express EGFP in tectal neurons (Dana *et al.*, 2019; Sakaki *et al.*, 2020). Electroporation parameters were 300ms train duration, -40V, 1ms/pulse, and 200pulses/s. The tadpoles were screened for expression of EGFP in single neurons after 48 hours, and imaged 72 hours post-electroporation. To record stimulus-evoked calcium activity, farnesylated (ie. membrane localized) jGCaMP7s was co-expressed with a farnesylated version of the red fluorophore mCyRFP1 (Laviv *et al.*, 2016) using a plasmid containing a self-cleaving P2A (Kim *et al.*, 2011). The red fluorophore mCyRFP1 served as a bright, photostable space-filler for tracking dendritic morphology. Single-cell electroporation parameters for this plasmid were 1.1s train duration, -40V, 1ms/pulse, and 200pulses/s.

***In vivo* imaging of single neuron activity:** Stage 48 Albino *Xenopus laevis* tadpoles were bathed for 5 minutes in 4mM pancuronium to temporarily paralyze them immediately prior to imaging. The tadpoles were then placed in a custom imaging chamber (Sakaki *et al.*, 2020) that was perfused with oxygenated 0.1x Steinberg's solution for the duration of the experiment. The structure and activity of the single labeled neurons were imaged at fast rates using a custom designed acousto-optic deflector (AOD)-based random access multiphoton microscope (Sakaki *et al.*, 2020). Calcium dynamics throughout the dendritic arbors of individual brain neurons were detected as changes in jGCaMP7s fluorescence. Branch radii were derived from a 3D image stack of the volume containing the complete neuron prior to the recording of calcium data.

**Visual stimulation protocol:** Visual stimuli were presented to the eye contralateral to the imaged tectum using a projector. Stimuli were composed of full-field brief (50 ms) flashes of OFF or ON stimuli, presented as a series of 4 pseudo-randomly spaced OFF stimulus (presented between 8-12 seconds apart) on an ON background, followed by a transition shift from an ON background to an OFF background and then 4 ON stimulus on an OFF background.

**Estimation of proportion of tuned vs untuned synapses in a dendritic arbor:** The value was generated from a dataset of 6 jGCaMP7s expressing *Xenopus* tectal neurons receiving OFF visual stimuli that produce stimulus-evoked action potential outputs. The calcium activity of all terminal dendritic branches was recorded, and the ratio of active branches was estimated by taking the proportion of branches with localized stimulus-driven branch-tip activity compared to non-responsive branches.

## Mathematical modeling

**Nernst-Planck equation:** Our model for ionic electrodiffusion and voltage dynamics in the dendritic arbor couples the Nernst-Planck equation with an electrical model describing membrane voltage. The Nernst-Planck equation describes the motion of ionic species in a fluid driven by both thermal diffusion, and electrostatic forces. This equation is built from the first principle of mechanics, stating the conservation of ions (Nernst, 1888):

$$\frac{\partial c_i}{\partial t} = -\nabla \cdot j_i \text{ for all ions } i, \quad (1)$$

where $c_i$ is the concentration of ion *i* in [mol.m$^{-3}$]. The flux of charge j$_i$ in [mol.s$^{-1}$.m$^{-2}$] sums the diffusion flux of ionic species derived from Fick's law $j_{diff}^i = -D_i \nabla c_i$, and an electric flux representing the motion of ions due to the electric field: $j_{elec}^i = -D_i \frac{z_i e c_i}{k_b T} \nabla V$ (Nernst, 1888; Kirby, 2010). In this equation, $V$ represents the voltage, and $\frac{D_i z_i e c_i}{k_b T}$ the Einstein relation for the mobility of ion species: $D_i$ is the diffusion coefficient of species i, $z_i$ the valence, $k_B T$ the thermal energy and $e$ is the electric charge (Kirby, 2010). We get:

$$j_i = j_{diff}^i + j_{elec}^i = -D_i \nabla c_i - D_i \frac{z_i e c_i}{k_b T} \nabla V. \quad (2)$$

A dendritic arbor is a complex tree-like structure where different compartments have various radii (Fig. 1A). We model this neuronal geometry by setting each dendritic segment as a cylinder with a specific radius (Fig. 1B). We consider the integrated flux along one branch of the tree of radius $a$, $J_i = \pi a^2 j_i$ in [mol.s$^{-1}$], which gives a one-dimensional ionic flux in space (Fig. 1B).

To add the contribution of ionic fluxes coming from membrane channels, we consider $j^{m,i}$, the flow per unit surface of species $i$ coming from the channels (in [mol.s$^{-1}$.m$^{-2}$]). Considering the radius $a$ of the cylindrical branch, the integrated flux over the membrane becomes $J^{m,i} = 2\pi a j^{m,i}$ (Fig. 1B). Finally, eq. (1) becomes:

$$\pi a^2 \frac{\partial c_i}{\partial t} = -\nabla \cdot J_i + J^{m,i} \quad \text{for all ions } i. \quad (3)$$

**Chemical reactions:** We consider general chemical reactions in the cytosol, taking the form:

$$\sum_{i=1}^{n} \alpha_i A_i \Leftrightarrow \sum_{i=1}^{m} \beta_i B_i \quad (4)$$

between ionic species $A_i$, $i = 1, ..., n$ and $B_i$, $i = 1, .., m$, with forward and backward reaction rates $k_f$ and $k_b$ respectively. The forward and backward reaction speeds are respectively:

$$v^f(x,t) = k_f \prod_{i=1}^{n} (c^{A_i}(x,t))^{\alpha_i}$$

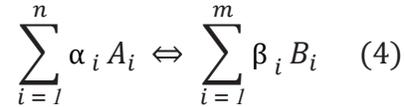

$$v^b(x,t) = k_b \prod_{i=1}^{n} (c^{B_i}(x,t))^{\beta_i}$$

where $c^K$ represents the concentration of species K. Hence, the variation of concentration of each species, due to the reaction at position $x$ and at time $t$ is:

$$\gamma_{A_i}(x,t) = \alpha_i(v^b(x,t) - v^f(x,t))$$

for the reactants $A_i$, and

$$\gamma_{B_I}(x,t) = \beta_i(v^f(x,t) - v^b(x,t))$$

for the products $B_i$. The final equation for the variation of concentration of each ionic species is given by:

$$\pi a^2 \frac{\partial c_i}{\partial t} = -\nabla \cdot J_i + J^{m,i} + \pi a^2 \gamma_i \quad \text{for all ions } i. \quad (5)$$

**Voltage dynamics:** The electric flux $J^i_{elec}$ of species $i$ in the Nernst-Planck equation (eq. (1)), requires the voltage profile along the branches. To build an equation for the voltage, we assume that the neuronal membrane behaves as a capacitance, i.e. the total charge density $q$ (in [C.m$^{-2}$]) at the membrane is proportional to the voltage difference $V$ (in [V]) across the membrane, with proportionality coefficient $c_{cap}$ (in [F.m$^{-2}$]):

$$q = c_{cap} V \quad (6).$$

Integrating over the membrane, we get the lineic charge $Q = 2\pi a\, q$, in [C.m$^{-1}$] (Fig. 1**B**). Using that the center of the branch is electroneutral (Stinchcombe *et al.*, 2016), and hence the charges are all located at the membrane, the charge $Q$ at the membrane is equal to the integrated charge over the section of the branch: $Q = \pi a^2 \sum_{i\,ions} c_i z_i N_a e$ where $N_a$ is the Avogadro number. Deriving eq. (6) over time, we obtain using the expression for $Q$:

$$c_{cap} \frac{\partial V}{\partial t} = \frac{a}{2} \sum_{i\,ions} \frac{\partial c_i}{\partial t} z_i N_a e. \quad (7)$$

The system of equations given by eqs. (1,2) and (5) for all ions, coupled to eq. (7) defines our model for voltage and ionic electrodiffusion in dendritic branches. The equations defining the channels dynamics $J^{m,i}$ are described below.

**Membrane channels dynamics:** To model the influx of ions through voltage-gated channels, we use a simplified Hodgkin-Huxley model for the sodium, potassium and leak channels (Hodgkin and Huxley, 1952), to which we add a Hodgkin-Huxley type channel for calcium current (Guerrier and Holcman, 2017). The classical Hodgkin-Huxley dynamics for the sodium, potassium and leak currents is reduced to a 2-dimensional system, using the approximation $h = (0.89 - 1.1n)$, and $m = m_{inf}$ (Izhikevich, 2007):

$$\frac{dn}{dt} = 0.1\, \alpha_n (1-n) - \beta_n n$$

$$I_{Na} = g_{Na}\, m_{inf}^3 h (V - E_{Na})$$

$$I_K = g_K n^4 (V - E_K)$$

$$I_L = g_L (V - E_L)$$

$$m_{inf} = \frac{\alpha_m}{\alpha_m + \beta_m}$$

and for k = n,m:

$$\alpha_k = \frac{1}{\tau_k} \frac{\theta_k - V}{\exp\left(\frac{\theta_k - V}{\tau_k}\right) - 1},$$

$$\beta_k = \eta_k \exp\left(\frac{-V}{\sigma_k}\right), \text{ for k = n, m.}$$

The membrane resting potential of the cell is 0 mV.
The calcium current dynamics is given by (Guerrier and Holcman, 2017):

$$I_{Ca} = g_{Ca}\, p^3\, l (V - E_{Ca})$$

and for $\varkappa$ = p,l:

$$\frac{d\varkappa}{dt} = \frac{1}{\tau_\varkappa} \left( \frac{1}{1 + exp(\theta_\varkappa - V)} - \varkappa \right).$$

The sodium, potassium and calcium currents are ion specific, and thus can be easily converted into ionic fluxes through the formula $J^{m,i} = \frac{I_i}{N_a z_i e}$. To maintain coherence within the Hodgkin-Huxley model, we consider that the leak current is only driven by chloride.

We included N-methyl D-aspartate (NMDA) receptors in our simulations, to mimic excitatory postsynaptic potentials. The mean approximation for the current entering the domain is (Koch, 1999):

$$I_N(t) = g_N \frac{e^{-t/\tau_{N,1}} - e^{-t/\tau_{N,2}}}{1 + 0.33[Mg2+]e^{-0.06(V-65)}} (V - E_N).$$

where we use the NMDA receptors conductance $g_N = 0.2$ nS and Nernst potential $E_N = 0$. To estimate the total amount of calcium ions entering through the NMDA receptors, we use that the fraction of the current carried by calcium ions is 15%. All parameters are given in Table 1.

**Decoupled model:** We decoupled the system of equations (1)-(5) and (7), using two assumptions: the first assumption is that the total diffusive flux is negligible compared to the total electric flux: $\sum_i z_i J^i_{diff} \ll \sum_i z_i J^i_{elec}$. The second assumption is that the electric conductivity of the cytoplasm $\sigma = \sum_{i\,ions} N_a \frac{D_i z_i^2 e^2 c_i}{k_b T}$ is constant (in [S.m$^{-1}$]). Using these two assumptions combined with eq. (1,2), we recover that the electric current along one branch is proportional to the gradient of the voltage: $I_R = N_a e \sum_{i\,ions} z_i J_i = \pi a^2 \sigma \nabla V$. We finally get the standard cable equation for the voltage:

$$c_{cap} \frac{\partial V}{\partial t} = I_m + \frac{a}{2} \sigma \frac{\partial^2 V}{\partial x^2}, \quad (8)$$

where $2\pi a\, I_m = N_a\, e \sum_{i\,ions} z_i J^{i,m}$.

Hence, the strategy of the decoupled model is to first compute the voltage using eq. (8), and then plug these dynamics into the Nernst-Planck equations eq. (1)-(2), and solve the ordinary differential equation for the ionic species we are interested in. Simulations of this simplified model are realized with the *Sinaps* python library the authors developed previously (Galtier and Guerrier, 2022).

## Data analysis

**Data filtering procedures - data normalization:** We observe a substantial variability in the fluorescence data between different nodes before the visual stimulation, during which the neuron is not receiving stimulus-evoked synaptic inputs. To explain this variability, we note that the fluorescence data reflects the number of photons received in a fixed period of time, and hence is representative of the number of jGCaMP7s-Ca present at the focal point of the laser position. For a given concentration $c_0$ of jGCaMP7s-Ca, in a dendrite of radius $a$, the number of observed photons will be proportional to $\pi a^2 c_0$. Hence, the fluorescence can vary with the arbor radii. This explains why the soma in the experimental data is very bright compared to the rest of the arbor, and why small filopodia are typically dim. To estimate the concentration of jGCaMP7s-Ca, we normalized each trace so that the mean fluorescences observed during the time before the first visual stimuli are the same. The fluorescence data were normalized only at this stage, to maintain a similar level of noise in each trace.

**Data filtering procedures - denoising:** To remove noise in the fluorescence data, we applied a rolling window filter to the raw data. We used a gaussian window, defined as $w(n) = e^{-n^2/2\sigma^2}$, with the window size n = 50 data points and the standard deviation parameter $\sigma = 10$.

**Synapse localization algorithm:** To localize synaptic activity, we designed an algorithm based on the rising dynamics of jGCaMP7s-Ca. We first applied our filtering procedure to generate a smooth curve representing the data as described above. At each recorded spatial location, we obtained an estimate of the rising slope of the time-dependent signal by taking the derivative in time at 60 ms following the visual stimuli. We then detected local maxima of the rising slope in space, which gave us the potential active synapses locations. A local maximum is defined as a point at the max over its five neighboring points.

# Results

**Simulation of the coupled model:** We implemented the coupled model given by eq. (1)-(5) and (7) (Method section, Fig. 2), in a simple branch attached to a soma. At the soma, we set up a Hodgkin-Huxley dynamics with sodium, potassium, and leak, as well as voltage-gated calcium channels (Method section). The soma radius and length are 2.5 and 5 µm respectively, and the branch radius and length are 1 and 50 µm. The ion flow throughout the branch is passive. We applied a current of amplitude 7 pA lasting 0.25 ms at the soma, to trigger an action potential. We plotted the potential and calcium dynamics, as well as the electrical conductivity, and the calcium diffusive and electric currents (Fig. 2, solid lines). We observe, as expected, that the action potential propagates very quickly within the branch (Fig. 2**A**, solid lines). We also observe a rise in calcium concentration due to the entry of calcium through voltage-gated calcium channels at the soma. This calcium propagates inside the branch due to the diffusive and electric fluxes (Fig. 2**B**, solid lines). To understand the contribution of both fluxes to calcium dynamics, we plotted $J_{elec}^{Ca}$ and $J_{diff}^{Ca}$, in the soma and at one position inside the branch (Fig. 2**C**). The maximum of the electric calcium flux is around $1.5\ 10^{-3}$ µmol.µm$^{-2}$.ms$^{-1}$ in both the soma and the branch, whereas the maximum of the diffusive flux is around $5.7\ 10^{-5}$ µmol.µm$^{-2}$.ms$^{-1}$ in the soma and drops to $1.7\ 10^{-7}$ µmol.µm$^{-2}$.ms$^{-1}$ in the branch. We also observe that the electric flux is on a faster time scale compared to the diffusive fluxes. These properties are maintained for sodium, potassium, and chloride ions. This shows that the ionic dynamics in this framework are driven by electric fluxes, and support the hypothesis used in the decoupled model: $\sum_i z_i J_{diff}^i \ll \sum_i z_i J_{elec}^i$. We also plotted the variations of the electrical conductivity of the cytosol $\sigma = \sum_{ions} \frac{D_i z_i e\ c_i}{k_b T}$ in Fig. 2**D**. We observed a maximal variation of up to 0.05 % in $\sigma$, which is coherent with the second assumption of the decoupled model, that $\sigma$ is constant.

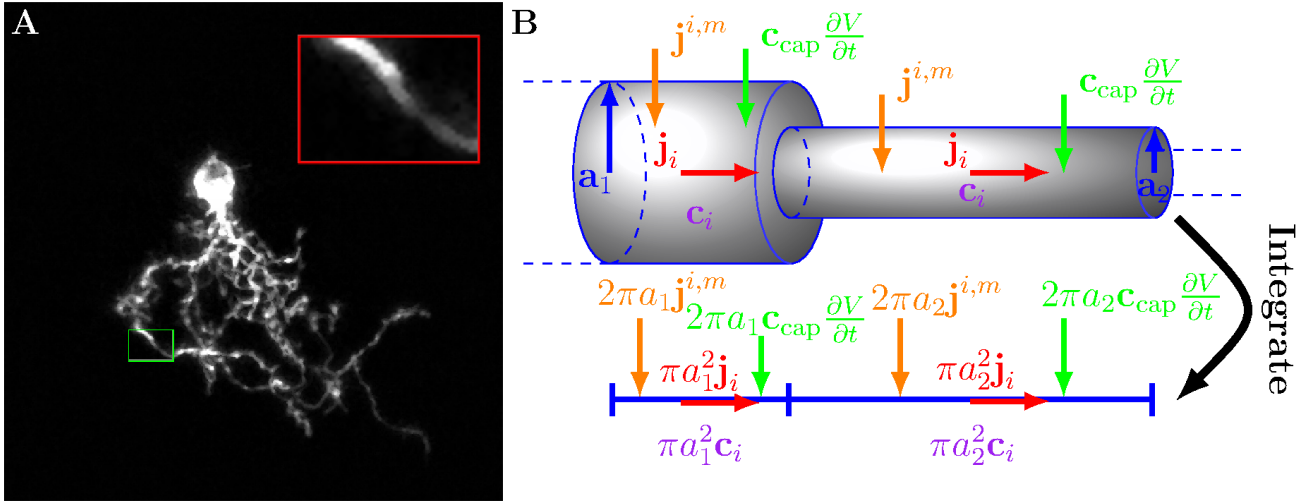

Figure 1: **A one-dimensional model of a three-dimensional dendritic arbor**. **A:** Z-stack of a *Xenopus* tectal neuron expressing EGFP, imaged while the animal was awake. Inset image is a 4.0x zoom of a section of dendrite with progressive thinning modeled in the panel to the right. **B:** Schematic representation of our modeling approach for two nodes of a dendritic branch analogous to the inset in A. The entire dendritic arbor is represented as a tree structure containing many nodes, where each branch of the tree is represented by multiple three-dimensional cylinders. Integrating over the cylindrical geometry, the model is reduced to a one-dimensional tree structure.

**The coupled vs decoupled model:** The coupled model can be simplified by decoupling the two equations which results in faster simulation. Our strategy is to first compute the voltage dynamic inside the neurons (eq. (8)), and plug this voltage into the calcium concentration equation to recover the dynamics of calcium (eq. (1,2), see Method section for further details). We show in Fig. 2 the difference in calcium and voltage dynamics in the coupled model (solid lines) versus decoupled model (dashed lines). We observe no differences between the two curves in voltage dynamics, and a small difference of less than 4 % in calcium dynamics. The differences in sodium and potassium dynamics are less than 1 %. To simulate voltage dynamics and ionic electrodiffusion in a full dendritic arbor geometry, we will utilize the decoupled model.

In both the coupled and decoupled models, we keep track of all the ionic species, using the standard ionic concentrations as has been previously described (Alberts, 2002). Due to the absence of sodium, potassium, and calcium reuptake, which all take place at a slower time scale, the concentrations of the different ions do not return to their resting values after an action potential. Therefore, the change in the electric conductivity $\sigma$ past 20 ms is due to an imprecision of our model at long time scales. To fully describe the variation of ionic concentrations in the cytosol in a longer time scale, it is necessary to add to this framework a model for the $Na^+/K^+$ ATPase exchanger (Øyehaug *et al.*, 2012), as well as for calcium extrusion, such as NCX or PMCA.

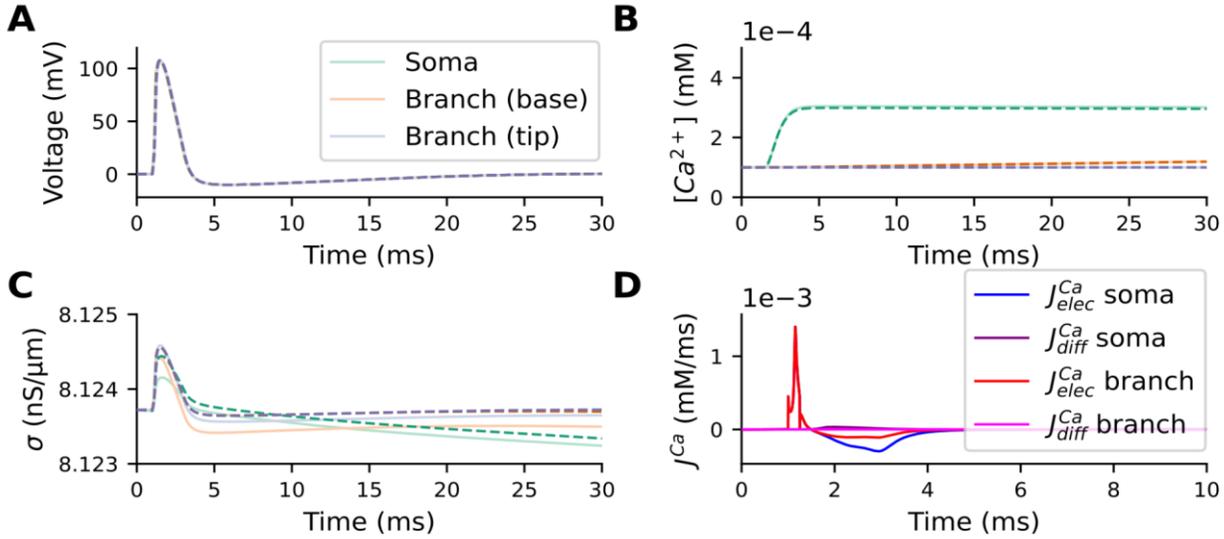

Figure 2: **Comparison between the coupled and the decoupled model:** Simulation of ionic dynamics in an active soma, connected to a passive branch. An applied current triggers an AP in the soma that propagates to the branch. Comparison between the coupled (solid lines) and the decoupled (dashed lines) models, at three different sites: the soma, and the base and tip of the branch. Time dynamics of voltage (all lines overlap) (**A**), calcium concentration (**B**) and electrical conductivity σ (**C**) at the soma and along the branch. **D:** Time dynamics of the currents $J^{Ca}_{elec} = \pi\, a^2\, j^{Ca}_{elec}$ (blue and red) and $J^{Ca}_{diff} = \pi\, a^2\, j^{Ca}_{diff}$ (purple and pink) at the soma, and in the dendritic branch 5 µm from the soma.

**Simulation of calcium and jGCaMP7s dynamics in a dendritic arbor:** Our goal is to realize numerical simulations recapitulating the fluorescence recordings obtained in the dendritic arbor of brain neurons recorded from *Xenopus laevis* tadpoles *in vivo*. Awake tadpoles expressing a membrane-localized jGCaMP7s in individual brain neurons were immobilized and visual stimuli were presented while full-neuron morphology and fast imaging of calcium dynamics were recorded using a random-access multiphoton microscope (Sakaki *et al.*, 2020).

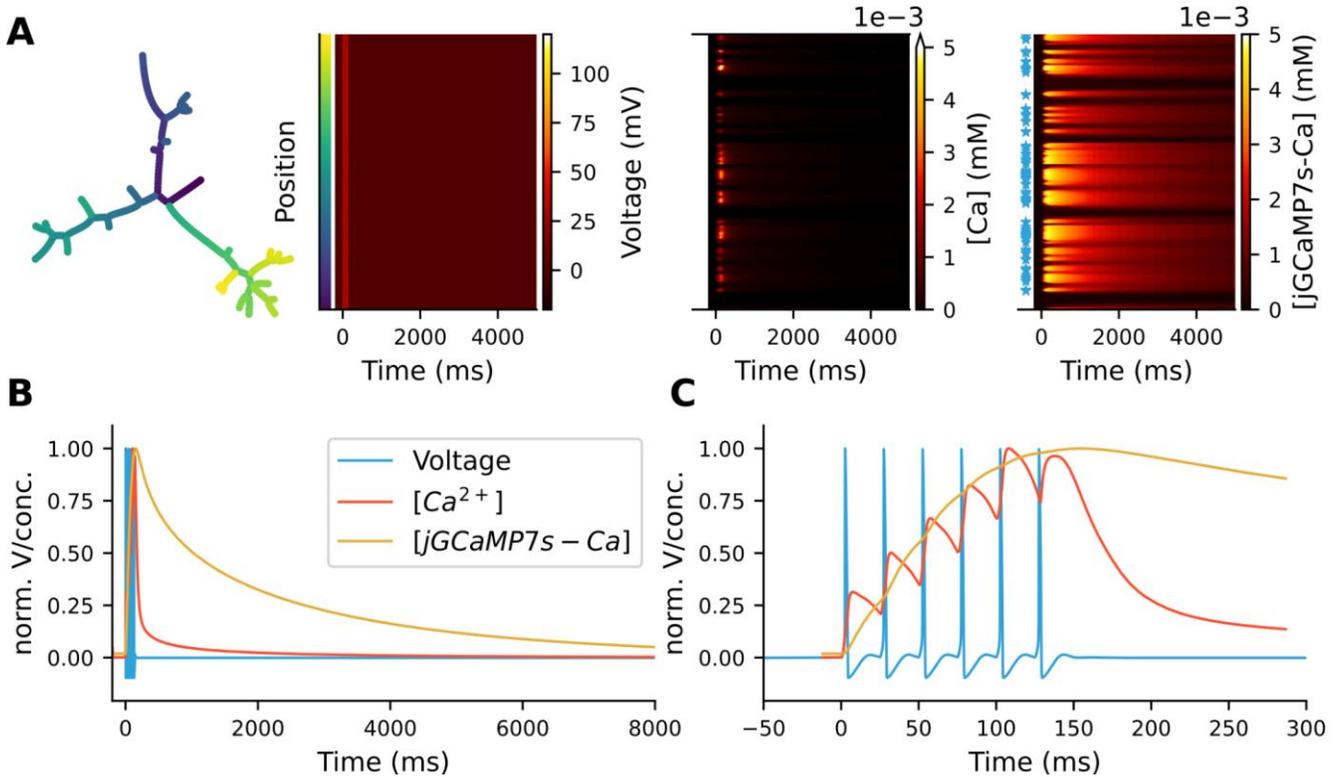

Figure 3: **Simulation of voltage, calcium and jGCaMP7s-Ca dynamics in a dendritic arbor. A:** Left: Tree-like representation of a simulated neuron, derived from the morphology of a real neuron expressing jGCaMP7s. Middle and Right: Comparison between the time scales of voltage, calcium, and jGCaMP7s-Ca concentration dynamics in simulated dataset, with the position of recorded activity indicated by the colors on the y-axis corresponding to color labels on the neuronal structure. The blue stars on the y-axis in the jGCaMP7s-Ca field plot represent synapses locations. **B:** Time dynamics of normalized voltage, calcium, and jGCaMP7s-Ca concentration. **C:** Magnification of the traces in **B**.

We imported the full dendritic geometry of a neuron using our Python library, *Sinaps* (Fig. 3**A**). Our dataset includes the radii, ranging from 1µm (across the dendrites) to 8.5 µm (across the soma), with a mean of *2.8 ± 2.5* µm. The dendritic geometry is represented as a tree structure, composed of nodes every 2 µm, which gives a total of 268 nodes. The terminal portion of a branch (i.e. not connected to a new branch) is called a leaf. We then modeled the influx of ions into the dendritic tree through different channels: to represent active channels, we implemented Hodgkin-Huxley type channels with potassium, sodium, calcium, and leak. We set Hodgkin-Huxley channels to be distributed everywhere on the surface of the dendritic arbor. At the soma, we set Hodgkin-Huxley channels with a conductance multiplied by 3. This was informed by previous modeling replicating experimental results that indicate that the soma has a higher density of voltage gated ionic channels (Bono and Clopath, 2017). To mimic synaptic activity, we added NMDA receptors to several nodes (see complete description in Method) (Wu *et al.*, 1996; Li *et al.*, 2011). To determine which node will be given a synapse, we used the results on synaptic repartition given in (Li *et al.*, 2011). First, Li *et al.* (2011) observed that 93% of terminal dendritic branches are receiving synaptic contact. To account for this proportion, we randomly chose 22 leaves over the 24 of our tree and added a synapse at those leaves. Second, the average synapse density reported in (Li *et al.*, 2011) is 0.43 synapses.µm$^{-1}$. The dendritic arbor we are considering is 534 µm long, which corresponds to approximately 229 synapses. Considering that we already set 22 synapses, we randomly distribute the 207 remaining synapses along the tree. Finally, we account for the fact that even in neurons that produce action potentials in response to a stimulus, only a subset of synapses on that neuron is going to tuned to be responsive to that particular stimulus by setting, for each synapse, a probability of being tuned to a stimulus of 0.24 (see Methods).

Calcium dynamics inside the cytosol are then coupled to the jGCaMP7s dynamics through the chemical reaction:

$$Ca^{2+} + jGCaMP7s \Leftrightarrow jGCaMP7s - Ca \quad (9)$$

with association rate $k_f = 21.5$ mM$^{-1}$.ms$^{-1}$ and dissociation rate $k_b = 0.00286$ ms$^{-1}$ (Dana *et al.*, 2019). We model calcium extrusion from the cell using a simple reaction Ca$-> \emptyset$ with reaction rate $k_{ex} = 0.03$ ms$^{-1}$. The initial calcium concentration in the cytosol is $c^{Ca} = 10^{-4}$ mM, and the jGCaMP7s concentration is $c^{jGCaMP7s} = 5. \, 10^{-3}$ mM. We performed simulations for different values of the jGCaMP7s concentration, ranging from $10^{-3}$ mM, to $10^{-2}$ mM, and observed no qualitative changes in the results. Simulations with the jCGaMP7s initial concentration $c^{jGCaMP7s}$ around $10^{-3}$ mM and below resulted in buffer saturation throughout the arbor, which is not consistent with experimental results. In simulations with $c^{jGCaMP7s}$ above $10^{-2}$ mM, we observe that jGCaMP7s is in excess, which likewise does not match with experimental observations. As the jGCaMP7s and jGCaMP7s-Ca employed here are membrane-bound (see Methods), we assume that their diffusion in the membrane is negligible. We represent the arrival of an input coming from the visual pathway via retinal ganglion cells (RGC), as a train of 6 synaptic inputs at 40 Hz (Demas *et al.*, 2011; Honda *et al.*, 2011). Synaptic inputs are modeled through the opening of NMDA receptors at synapses.

We perform numerical simulations using the Python library *Sinaps* (Galtier and Guerrier, 2022), Fig. 3. From this, we observe a time scale difference between the action potential dynamics, that lasts for a few ms, to the calcium concentration dynamics, lasting a few hundred milliseconds, to the fluorescence calcium sensor dynamics lasting several seconds (Fig. 3**B-C**). Additionally, we observe that the slower time scale of jGCaMP7s-Ca dynamics 'filters' voltage and calcium signals, making it more difficult to infer calcium and voltage dynamics from jGCaMP7s-Ca data.

**jGCaMP7s-Ca rise dynamics are faster when proximate to synaptic activity:** The simulations of jGCaMP7s-Ca dynamics in the full dendritic arbor reveals the filtering effect of the change in time scale (Fig. 3**B**). Hence, the detailed dynamics of calcium entry is not easily observable in fluorescence data (Fig. 4**A**). We also observe that the region close to the soma has the highest stimulus-evoked fluorescence amplitude, which is due to the large influx of calcium ions at the soma. This shows that the maximum intensity of fluorescence over the tree is not a good marker of synaptic activity. We also observe, in the simulations, that the spread of calcium is contained around its entry points, due to the limited diffusion of calcium before extrusion. Our algorithm makes use of this local property to discriminate synaptic activity from other calcium events. Indeed, we observed in simulations typical patterns of synaptic activity (Fig. 4**B**), with a conical shape. These diffusion patterns start at synapses and propagate to neighboring locations. The typical diffusion length of those patterns, $l_{diff} = 10 \, \mu m$, was estimated from the simulations. After applying a denoising algorithm to the data (Fig. 4**A**), we observed these typical patterns in the experimental data as well (Fig. 4**B**). We also observed in simulations 'black regions', far from synapses (Fig. 4**B**). These black regions are also somewhat observed in experimental data (Fig. 4**A-B**). They are a result of the limited intracellular diffusion of calcium ions. The diffusion of calcium ions coupled to the sensor is negligible since jGCaMP7s is anchored to the plasma membrane. Using these observations, we developed an algorithm to detect synaptic activity in jGCaMP7s-Ca dynamics. The algorithm is based on identifying local spatial maxima of the rising slopes of jGCaMP7s-

Ca along the dendritic arbor. Indeed, in our simulations, we observed that the closer a location was to a synapse, the higher the rising slope of the jGCaMP7s-Ca dynamics (Fig. 4**C**). This result is robust and was observed for different synapses across multiple locations. Hence, to detect synaptic activity in the experimental data, we built an algorithm detecting local maxima in space of the rising slopes (Fig. 4**D**, see Method). Note that in Fig. 4**D**, the local maxima need to be considered along the arbor topology, and not linearly.

In the experimental protocol, four visual stimuli each composed of a 50ms OFF stimulus were applied. To test the robustness of our algorithm, and to investigate the variations of synaptic activity across different individual presentations of the visual stimulus, we applied our algorithm after each stimulus, and compared the localization of synapses (Fig. 4**D**). We display the result of our algorithm for two stimuli and observe that some synaptic sites are active in both stimuli, and others are active only for a single stimulus (Fig. 4**D-E**).

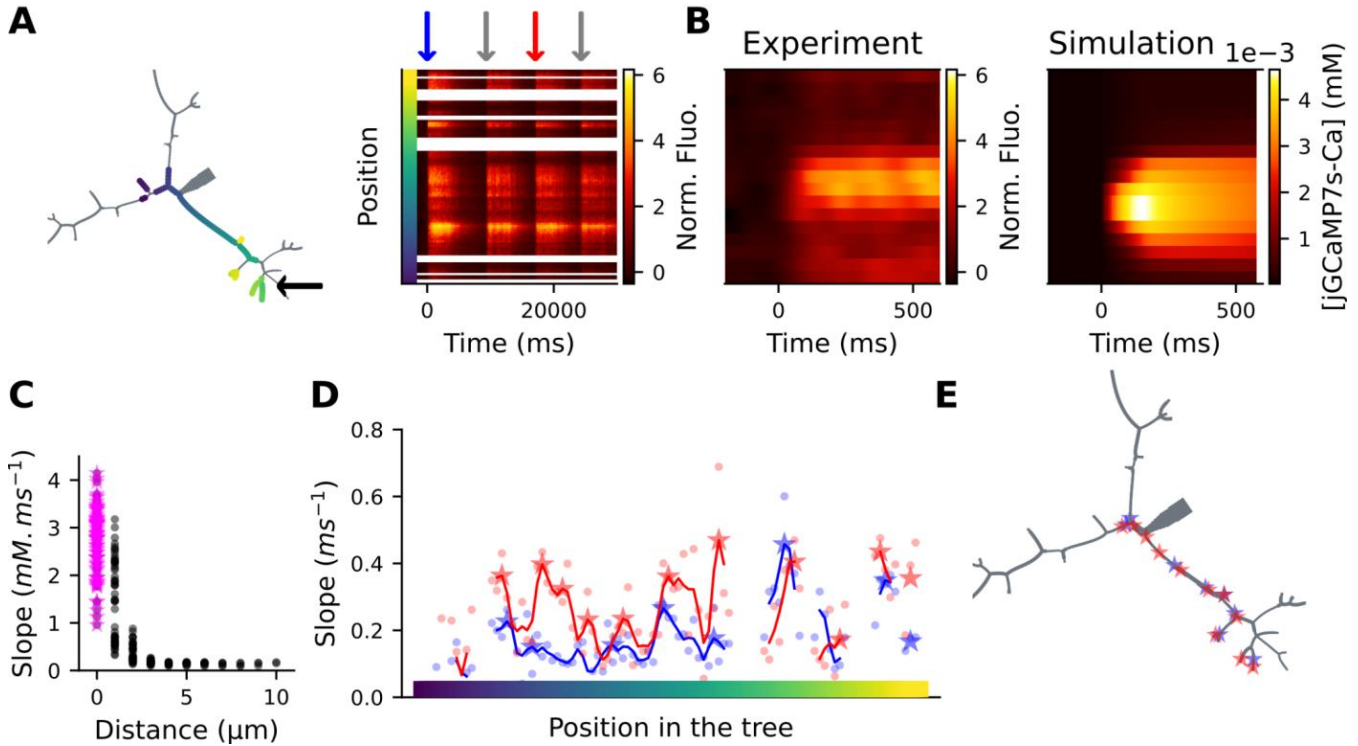

Figure 4: **The rising slopes of stimulus-evoked calcium transients indicate the location of active synapses in simulation and experimental data. A:** Left: structure of a dendritic arbor generated from *in vivo* imaging of a jGCaMP7s-expressing tectal neuron. Right: visual-evoked Ca responses of this neuron, with the position of recorded activity indicated by the colors on the y-axis corresponding to color labels on the neuronal structure. Stimulus times are indicated by the colored arrows on top of the field plot. **B:** Left: experimental recordings in the branch indicated with a black arrow in the tree structure in panel **A** processed through a Bessel filter, demonstrating a typical conical shape of visual-evoked discrete dendritic responses, consistent with simulations of synapses (Right). **C:** Rising slopes of the simulated jGCaMP7s-Ca concentration, plotted according to the distance from the closest synapses labeled with magenta stars. **D:** Plot of the rising slopes of Ca transients demonstrating how our algorithm identifies local maxima in space across a dendrite in the filtered and normalized fluorescence traces, for two visual stimuli (red and blue arrows in panel **A**). Dots represent the slopes at each point at t=60ms following visual stimuli, solid lines represent a fit smoothed in space (see Methods). The positions along the tree of the neuron in **A** are represented by the color bar in the x-axis. Stars indicate synapses detected by the algorithm. **E:** Dendritic arbor as in **A**, with stars at synaptic locations detected by our algorithm, for the two visual stimuli in **D** (red and blue, colors correspond).

**Identification of active synapses within the dendritic arbor:** To validate the algorithm presented above (Fig. 4), we ran simulations of the full dendritic arbor, with synapses at locations determined by the algorithm. Note that only a part of the dendritic arbor was imaged (Fig. 4**A**). For the rest of the tree with no simulations, we maintained our previous synaptic distribution and organization. As calcium dynamics at synapses is a local behavior, the position of synapses in the rest of the arbor does not influence the dynamics at recorded positions. We then compare the rising slopes between experimental data and simulations (Fig. 5**A**) and observe a very good agreement between them. We emphasize here that we normalized fluorescence data to homogenize the noise before the first stimulus (Method), and this normalization *a priori* is sufficient to generate a good correspondence between the fluorescence amplitude and jGCaMP7s-Ca simulated dynamics (Fig. 5). We observe that the decay slope in experiments is well replicated by our model in a subset of the nodes (Fig. 5B-C, green and red nodes). We also observe regions with an additional increase in fluorescence-based calcium signal (blue, orange, and purple nodes), that we hypothesized may be driven by local calcium-evoked release from intracellular calcium sources within the dendrite.

# Discussion

Technological advancements in imaging platforms now allow for recording neurons with unprecedented spatial and temporal resolution (Kazemipour *et al.*, 2019; Sakaki *et al.*, 2020). When combined with advances in calcium sensors (Dana *et al.,* 2019), it has become possible to sample across the complete dendritic arbor and soma of a neuron at rates enabling the detection of fluorescence signals driven by a wide range of calcium sources including synaptic potentials, back-propagating action potentials, voltage-gated calcium channels and endoplasmic reticulum-mediated calcium release. These new datasets raise the challenge of developing methods to segregate these sources for independent analyses. Here, we have created a model to predict synaptic activity and their locations on dendritic arbors from these complex datasets.

**An accurate model for calcium, jGCaMP7s and jGCaMP7s-Ca dynamics:** We generated a model for voltage propagation and ionic electrodiffusion in the dendritic arbor at the microscale level. This model is based on the Nernst-Planck equation for ionic electrodiffusion, ensuring the precision necessary to reproduce microscale level dynamics. We created a decoupled version of our model, allowing fast simulation in detailed dendritic arbor geometries achieved from *in vivo* morphometric imaging experiments. To derive the decoupled version, we show that the longitudinal ionic flux can be well approximated by a resistive current (j = $\sigma$ E). Our study also emphasizes the role of electrodiffusion in calcium propagation, versus simple diffusion. Hence, in Fig. 2**D**, we observe that the major part of the calcium flux is the electric flux $J^{Ca}_{elec}$, that is also faster than the diffusive flux $J^{Ca}_{diff}$.

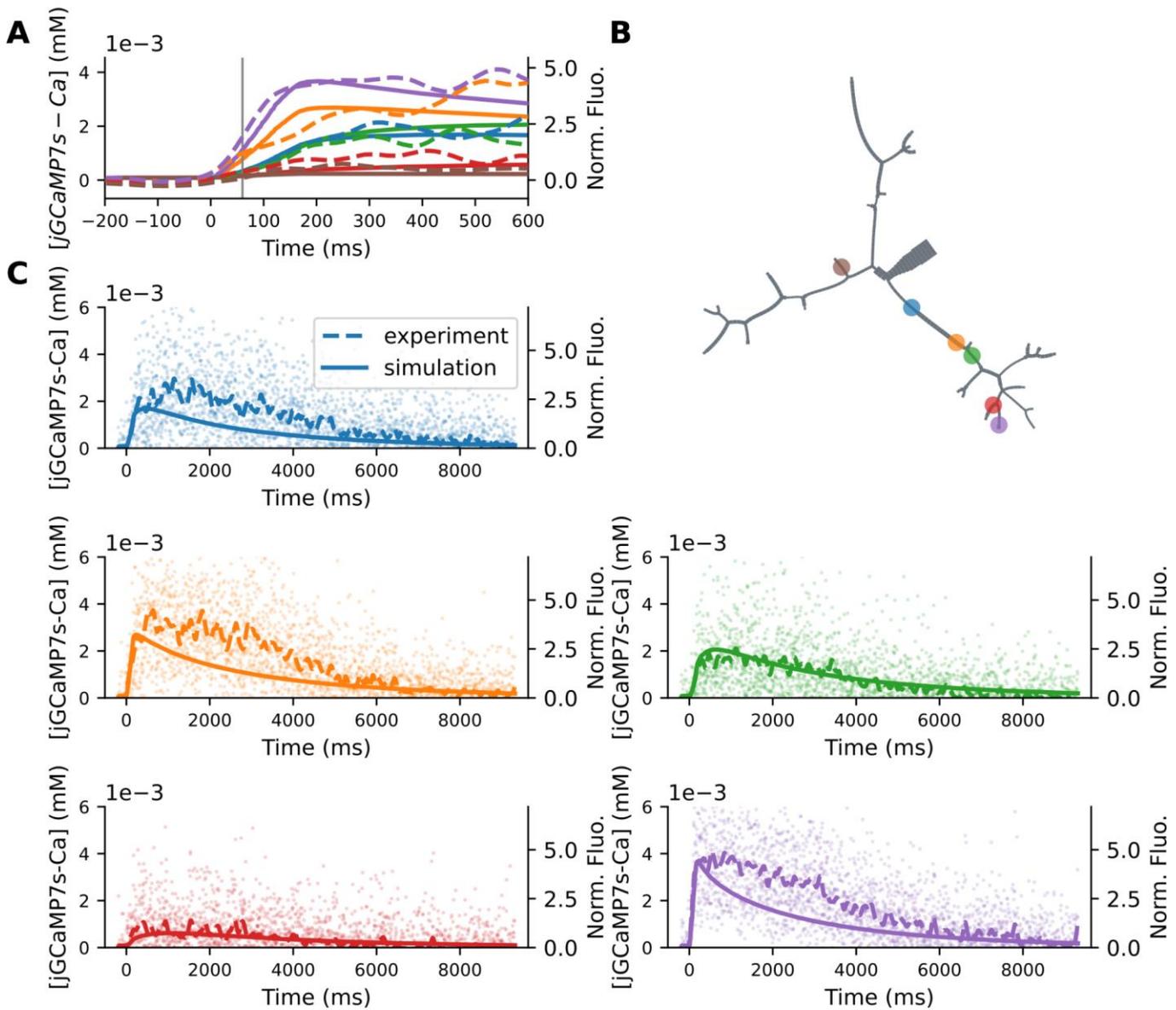

Figure 5: **Comparison between experimental and simulated calcium traces, with synaptic activity determined using our detection algorithm. A:** Comparison between the filtered fluorescence data (dashed lines, normalized) and simulations of jGCaMP7s-Ca dynamics (solid lines) at several locations in the tree. The gray bar represents the time at which the algorithm compares the rising slopes. **B:** Positions on the dendritic arbor of the nodes represented in **A** and **C** (colors correspond). **C:** Comparison between the raw fluorescence data (points), the filtered data (dashed lines) and the simulated jGCaMP7s concentration (solid lines) in 5 of the 6 nodes presented in **A** (colors correspond).

**Stimulus-evoked jGCaMP7s-Ca fluorescence rising slopes are a marker for synaptic activity:** Here, we propose an algorithm that uses the rising slope of stimulus-evoked calcium fluorescence recordings as a marker for synaptic activity. We observe from our simulations that the rising slope of the stimulus-driven jGCaMP7s-Ca influx is strongly dependent on the distance from active synaptic sites, and we employ this measure for determining the locations of active synapses in real datasets that have a high temporal sampling rate (Fig. 4). In our jGCaMP7s-Ca experimental recordings we were able to identify predicted active synapse locations by identifying spatial maxima of rising slopes along the dendritic arbor, and the activity at these sites was consistent with recorded retino-tectal synaptic activity (Engert *et al.*, Demas *et al.*, 2011; Honda *et al.*, 2011). We observe a strong agreement between our model's predictions of jGCaMP7s-Ca dynamics and fluorescence-based experimental measurements (Fig. 5). Indeed, a peak following the stimulus is observed in both the fluorescence traces and the jGCaMP7s-Ca concentration dynamics, and the rising slope and peak amplitude correspond in most nodes (Fig. 5**A** and **C**).

The differences in the rising slopes of stimulus-induced calcium signals (Fig. 5) are due to the different dynamics of the various calcium sources. In our simulations, NMDA receptor channels were added to model calcium entry at synapses, as well as voltage-gated calcium channels (VGCC), to represent calcium entry consequent to a sufficiently high membrane depolarization. The simulation yields stereotypical conical patterns of spatially and temporarily isolated calcium signals (Fig. 4**B**), characteristic of synaptic activity. This shape is due to a competition between the diffusion of calcium ions entering at a synapse, and free calcium removal from the cytoplasm, through binding on various buffers and extrusion through pumps. This spatial discrimination of the calcium signal is also due to our fluorescent jGCaMP7s sensor being membrane-bound and exhibiting minimal diffusion. Finally, these patterns are also easily observable due to the higher calcium current amplitude at active synapses via NMDA receptors, compared to regions with no active synapses, where the main calcium current is due to VGCCs in our simulations. In the experimental results, despite the inherent noise and additional calcium sources, we observe a similar pattern of stimulus-evoked calcium influx that we interpret as markers of synaptic activity. To detect these conical shapes, our algorithm searches for local maxima in space of the rising slopes. At other locations in the dendritic arbor we also observe calcium patterns that do not appear in simulations (Fig. 5, blue, orange and purple nodes). We hypothesize that the observed prolonged stimulus-evoked calcium transients may result from local calcium-evoked release of calcium from intracellular stores that is not not currently modeled (Segal and Korkotian, 2014).

**Utility of detecting synaptic activity from fluorescence Ca-sensor recordings:** Our algorithm detects and localizes synaptic activity from fluorescence recordings, without the need for synaptic markers such as fluorescently tagged synaptic proteins or tagged intrabodies that target these proteins (Gross *et al*., 2013). There is substantial utility for a model that can identify active synapses from fluorescence-based calcium data, since expression of synaptic markers or immunostaining to localize synapses is often not feasible for *in vivo* experiments. One challenge arises from the limited number of detection channels typically available in most multi-photon imaging systems. As a result, only a small number of different fluorescent markers can be imaged concurrently, such as a calcium sensor reporting activity in one channel and a neuronal morphology structural marker in the second channel. Furthermore, even in experiments pairing a synaptic marker with an activity marker, our algorithm would prove useful for discriminating synaptic responses from signals arising other sources.

**Limitations:** One limitation of our model for discriminating synaptic activity from fluorescence-based calcium signals is the requirement that the datasets have a high temporal resolution, since synaptic activity is discriminated based on the faster rise of the slope compared to other calcium sources. We estimate that to be able to accurately identify synaptic activity from jGCaMP7s fluorescence, recordings need to be performed at a rate of approximately 50Hz. Another limitation is the lack of exact experimental measurements for certain biological variables that by their nature are extremely difficult to quantify. For example, it would be extremely challenging to obtain the exact concentration of jGCaMPs7s expressed in neurons *in vivo*, despite the availability of previous research providing estimations of plasmid expression based on the promoter (Dou *et al*., 2020) due to likely variable amounts of plasmid delivered to each neuron. Consequently, we performed simulations for a range of different concentrations of jGCaMP7s, which indicate that the results were not significantly impacted.

**Future development of the algorithm:** The algorithm presented here is a first attempt to localize synaptic activity from fluorescence data in the full dendritic arbor of a neuron recorded in the intact and awake developing vertebrate brain. The resultant model successfully localizes suspected synaptic activity in an *in vivo* dataset, however there are several potential improvements that can be made. Firstly, while the signal given by our dendritic recordings is spatio-temporal, we expect some correlation of the noise at a spatial level as well as at the time level, however our denoising algorithm is currently only temporal. To address this, we plan to add the spatial component using the model for electrodiffusion. Secondly, we intend to expand the model to be able to discriminate a wider variety of calcium signals in a neuron. In addition to stimulus driven synaptic calcium, neurons in a stimulus-response circuit also have calcium transients from other internal and external sources that have been demonstrated to be biologically relevant. For example, endoplasmic reticulum-based calcium transients are believed to potentially play a role in modulating synaptic plasticity (Segal and Korkotian, 2014), as are back-propagating action potentials (Waters *et al*., 2005).

**Application of the algorithm to analysis of synaptic inputs:** This algorithm allows for the identification of synaptic inputs in neurons based on evoked calcium. A natural next step is to adapt it to identify synaptic and extrasynaptic inputs based on neurotransmitter input. This is a new possibility in the field due to the development of increasingly sophisticated fluorescence-based neurotransmitter sensors. In particular, the third generation of iGluSnFR sensors has recently been designed specifically to allow for the temporal discrimination of synaptic glutamate from extrasynaptic glutamate (Aggarwal *et al*., 2022).

**Building an integrated input-output neural model:** Subsequent to the adaptation of the model for the analysis and discrimination of synaptic glutamate, we intend to apply the model to the analysis of experiments in which both a glutamate sensor and a calcium sensor are simultaneously expressed in a single neuron *in vivo*. Thus, we can build a comprehensive model of both neurotransmitter input and the resultant calcium output and investigate input-output relationships at synapses.

# Conclusion

In this paper, we present a model and algorithm designed to detect possible synaptic activity in *in vivo* fluorescence-based calcium recordings. Our model is based on the Nernst-Planck system of equations for electrodiffusion, coupled to a capacitive equation representing voltage dynamics at the membrane. We then decoupled our model to allow for fast simulations at the scale of morphometric experiments. Using a dendritic arbor structure derived from experimental data, we simulated realistic dynamics of calcium while it is bound to the jGCaMP7s sensor. Using these simulation results, we identified a typical conical shape of calcium diffusion following its entry at a synaptic site. This conical shaped calcium transient was present in experimental calcium imaging datasets, enabling it to be used as a marker for synaptic activity. We then built an algorithm to automatically detect this marker. Re-running our decoupled model with the synaptic sites detected by our algorithm, we observed a very good agreement between our numerical simulations and the experimental dataset. We also observed calcium transient patterns not yet identified by our algorithm potentially due to variable number and frequency of AP input at individual synapses, or local release of calcium from intracellular stores, which we intend to adapt the model to be able to discriminate between their sources. Our model and algorithm are tools capable of identifying synaptic activity across a dendritic arbor in *in vivo* fluorescence-based calcium recordings, at the microscale level.

Table 1: Simulation parameters for the channels dynamics

| Parameter | Description | Value | Reference |
|---|---|---|---|
| $g_{Na}$ | conductance of Na$^+$-current | 120 mS.cm$^{-2}$ | (Hodgkin and Huxley, 1952) |
| $g_{Na,soma}$ | conductance of Na$^+$-current | $3 \times 120$ mS.cm$^{-2}$ | (Hodgkin and Huxley, 1952; Bono and Clopath, 2017) |
| $E_{Na}$ | equilibrium potential of Na$^{2+}$-current | 115 mV | (Hodgkin and Huxley, 1952) |
| $g_K$ | conductance of K$^+$-current | 36 mS.cm$^{-2}$ | (Hodgkin and Huxley, 1952) |
| $g_{K,soma}$ | conductance of K$^+$-current | $3 \times 36$ mS.cm$^{-2}$ | (Hodgkin and Huxley, 1952; Bono and Clopath, 2017) |
| $E_K$ | equilibrium potential of K$^+$-current | `-12 mV` | (Hodgkin and Huxley, 1952) |
| $g_L$ | conductance of leak current | 0.3 mS.cm$^{-2}$ | (Hodgkin and Huxley, 1952) |
| $g_{L,soma}$ | conductance of leak current | $3 \times 0.3$ mS.cm$^{-2}$ | (Hodgkin and Huxley, 1952; Bono and Clopath, 2017) |
| $E_L$ | equilibrium potential of leak current | 10.6 mV | (Hodgkin and Huxley, 1952) |
| $g_{Ca}$ | conductance of Ca$^{2+}$-current | 14.5 mS.cm$^{-2}$ | (Guerrier and Holcman, 2017) |
| $E_{Ca}$ | equilibrium potential of Ca$^{2+}$-current | 115 mV | (Guerrier and Holcman, 2017) |
| $\tau_m$ | parameter for Na$^+$-current | 10 ms | (Hodgkin and Huxley, 1952) |
| $\theta_m$ | parameter for Na$^+$-current | 25 mV | (Hodgkin and Huxley, 1952) |
| $\eta_m$ | parameter for Na$^+$-current | 4 | (Hodgkin and Huxley, 1952) |
| $\sigma_m$ | parameter for Na$^+$-current | 18 | (Hodgkin and Huxley, 1952) |
| $\tau_n$ | parameter for K$^+$-current | 10 ms | (Hodgkin and Huxley, 1952) |
| $\theta_n$ | parameter for K$^+$-current | 10 mV | (Hodgkin and Huxley, 1952) |
| $\eta_n$ | parameter for K$^+$-current | 0.125 | (Hodgkin and Huxley, 1952) |
| $\sigma_n$ | parameter for K$^+$-current | 80 | (Hodgkin and Huxley, 1952) |
| $\tau_p$ | parameter for Ca$^{2+}$-current | 1.3 ms | (Guerrier and Holcman, 2017) |
| $\theta_p$ | parameter for Ca$^{2+}$-current | 102 mV | (Guerrier and Holcman, 2017) |
| $\tau_l$ | parameter for Ca$^{2+}$-current | 10 ms | (Guerrier and Holcman, 2017) |
| $\theta_l$ | parameter for Ca$^{2+}$-current | 24 mV | (Guerrier and Holcman, 2017) |
| $g_N$ | NMDA receptor conductance | 0.15 mS.cm$^{-2}$ | (Koch, 1999) |
| $E_N$ | NMDA receptor equilibrium potential | 75 mV | (Koch, 1999) |
| $\tau_{N,1}$ | NMDA receptor time constant | 11.5 ms | (Koch, 1999) |
| $\tau_{N,2}$ | NMDA receptor time constant | 0.67 ms | (Koch, 1999) |
| [Mg$^{2+}$] | Magnesium block | 2 mM | (Koch, 1999) |

# Abbreviations

N-methyl D-aspartate (NMDA)

Voltage-gated calcium channels (VGCC)

# Declarations

**Authors' contributions:** KH designed the project. CG and NG built the model and did the simulations. TDT performed experiments. CG, TDT and KH wrote the main manuscript text and NG prepared figures 1-5. All authors reviewed the manuscript.

**Competing interests:** The authors declare no conflict of interest.

**Ethics approval:** The animal study was reviewed and approved by the UBC Animal Care Committee and was in accordance with the Canadian Council on Animal Care (CCAC) guidelines. Animal Care Number: A190297

**Consent to participate:** *'Not applicable'*

**Data Availability statement:** The jupyter notebooks generating the different figures of the paper, as well as the experimental data presented in the paper are available at the following address: [PredictionSynapticActivity.zip](PredictionSynapticActivity.zip). The code is currently at the discretion of the reviewers, and will be made publicly available after publication on *Sinaps* documentation.

**Funding:** Financial support from CIHR grant #FDN-148468, AccelNet International Network for Bio-Inspired Computing and the Fyssen foundation is gratefully acknowledged.